\documentclass[doublecol,linenumbers]{epl2} 

\usepackage{amssymb}
\usepackage{color}

\title{Charge disorder and variations of $T_c$ in Zn-doped cuprate
superconductors}

\shorttitle{The variations of $T_c$ in Zn-doped cuprates} 

\author{E.V.L. de Mello
\and David M\"ockli
}
\shortauthor{E.V.L. de Mello \etal}

\institute{                    
Instituto de F\'{\i}sica, Universidade Federal Fluminense, 
Niter\'oi, RJ 24210-340, Brazil
}

\abstract{
Impurity doping like Zn atoms in cuprates were systematically studied
to provide important information on the pseudogap phase because
this process substantially reduces $T_c$ without effect on $T^*$. Despite many important
results and advances, the normal phase of these superconductors is still subject 
of a great debate. We show that the observed Zn-doped data
can be reproduced by constructing a nanoscale granular superconductor 
whose resistivity transition 
is achieved by Josephson coupling, what provides also a simple interpretation to the pseudogap phase.}

\begin{document}

\maketitle

\section{Introduction}

The effect of non-magnetic Zn impurity doping
substituting planar Cu was studied in several cuprate superconductors.
It was shown that it induces a local magnetic moment in the four neighboring Cu
sites by suppressing local antiferromagnetic correlation\cite{Alloul}
acting as a stronger carrier scatterer, which causes pair breaking
that substantially reduces\cite{Bernhard} $T_c$ and the onset of
Nernst signal $T_{on}$.\cite{Xu} On the other hand, Zn does not significantly modify
the pseudogap temperature $T^*$.\cite{Xu,Marta,Walker,Abe} 
In this paper we show that these results, even in different systems like
YBa$_2$Cu$_3$O$_{7-\delta}$ (YBCO) and La$_{2-x}$Sr$_x$CuO$_4$ (LASCO),
are consistent with a granular superconductor formed by electronic grains.

Some years ago scanning tunneling microscopy (STM) experiments detected nanoscale spatial variations in
the  electronic density of states in agreement with a granular structure.\cite{Lang,Gomes,McElroy,Pasupathy,Kato}
More recently, incommensurate electronic distributions have been measured
by many different experiments.
\cite{Wise,Ghiringhelli,Chang,Torchinsky,LeTacon,Comin,Neto} 
Resonant elastic X-ray scattering
(REXS) experiments in underdoped YBCO reported incommensurate scattering
vectors along the Cu-O planes and a corresponding temperature dependent correlation length.\cite{Ghiringhelli,Chang} 
Combined REXS and scanning tunneling microscopy (STM) on 
BSCCO found similar results but, with an increasing incommensurate
periodicity with doping and a larger onset temperature of $T\approx 300$K.\cite{Neto} 
Recently, TF-$\mu$SR experiments have also provided evidences for an 
intrinsic source of electronic inhomogeneity in
superconducting cuprates at temperatures much larger than $T_c$.\cite{Sonier:08,Mahyari}

To model this charge inhomogeneity we perform simulations of a nanoscale 
phase separation transition with an onset temperature $T_{PS}$ 
($\approx 300$K used here for YBCO) slightly 
above the pseudogap temperature $T^*$.\cite{Mello04,Mello09}
A natural way to simulate an electronic phase separation is through the 
Cahn-Hilliard (CH) differential equation.\cite{CH,Otton}
Solving numerically the CH equation it is possible to follow the 
time evolution of a conserved order parameter that yields
the local charge density distribution into nanoscale patches or puddles
of low and high local density. This scenario, originally proposed to phase separation of alloys, is in 
agreement with the observed charge inhomogeneities mentioned above, in particular
with the work of Ref. \cite{Wise}. We have also shown
that these patches occurs at the minima of the the free energy and are separated barriers that 
confine the charges, favoring the formation of 
"local" Cooper pairs at low temperatures.\cite{PhysC12}
On the other hand, 
we compare the Zn doped effect with that of 
site or bond dilution ferromagnet.\cite{Harris}
Comparing the effect of Zn concentration with the relative decrease of the average free
energy barrier between the patches, we can estimate the average effect of Zn doping
in the local $d$-wave superconducting gap, as observed before by STM for 
the case of Ni doping.\cite{Lang}
Using a simplified expression for Josephson coupling between the patches, we obtain $T_c$ as
function of Zn concentration in good agreement with the experimental
results.\cite{Xu,Marta,Walker,Abe}

\section{Simulations of Charge Inhomogeneities}

Following earlier indications of charge disorder in cuprates\cite{Muller}
and STM results on gap inhomogeneities\cite{Lang} we have proposed
an electronic phase separation described by the time dependent non-linear 
CH differential equation that breaks spatial invariance.
In this way we can
follow the time evolution of the local charge density as the 
conserved order parameter.\cite{Mello04,Mello09,Otton} 
The simulations show the formation of isolated domains
similar to a system quenched from high to low temperatures through
the phase separation transition. \cite{Bray}
This is a technical way to study a phase segregation
that leads to a nanoscale charge inhomogeneity similar to those
measured earlier,\cite{Muller} and also in recent experiments.\cite{Wise,Ghiringhelli,Chang,Torchinsky,LeTacon,Comin,Neto,Sonier:08,Mahyari}

The local order parameter yields the GL free-energy and we can follow 
the formation of free-energy ``wells'' surrounded by barriers,
with alternate regions of low and high density.\cite{PhysC12,EPL12a,EPL12b,EPL13} 
In this segregation process the charges loose part of their kinetic
energy what favors the formation of local Cooper pairs. This is the mechanism
by which cuprates form electronic granular superconductors, with
local pair amplitudes above $T_c$ without long range order.\cite{PhysC12} 

\begin{figure}[ht]
\begin{center}
\centerline{\includegraphics[width=6.5cm,angle=-90]{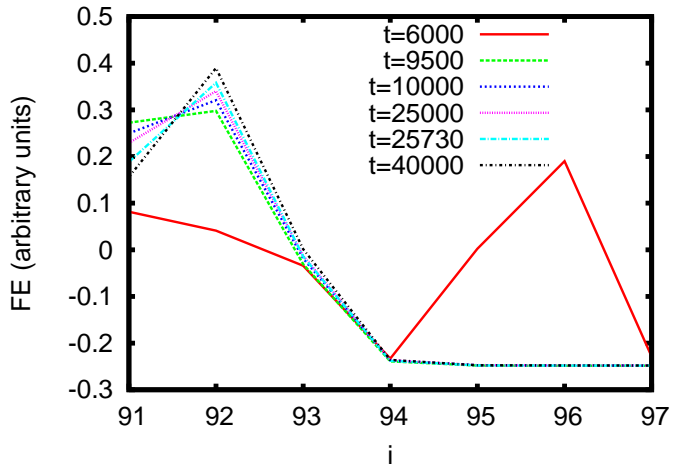}}
   \end{center}
\caption{(Color online)  Detail of the time evolution of the GL free-energy barrier between two patches along
a line in the x-direction.  t=40000 represents the ``infinity'' time phase separation or the system without Zn. 
The Zn doped system is simulated by lowering the potential barrier, as discussed
in the text. }
\label{FEtimeEvol} 
\end{figure}

This approach was successful to relate the time of phase separation boosted
by X-ray irradiation and the respective variations of $T_c$ in the 
La$_{2}$CuO$_{4-y}$ system.\cite{EPL12a} Here we adopt a similar approach 
and relate the presence of Zn concentration $x$ with
the specific time of phase separation. This is done in analogy with a dilute 
ferromagnet which the (site or bond) dilution is related with the 
Curie temperature or the average coupling.\cite{Harris} 
Then the Zn concentration is relate with the time evolution of 
the height of the free energy barrier between the patches or localization degree. 
In fact, since each Zn atom affects five Cu atoms, the one it substitutes and
the four neighbors, we relate the percentage $x$ of
Zn with a decrease of $5x$\% of the GL free-energy barriers. The
change of strength with the time evolution of a single barrier
is shown in Fig.(\ref{FEtimeEvol}). A similar figure involving the time evolution of two free-energy barriers between two patches is shown in Ref. \cite{EPL12a}.

\section{Calculations of Local Pairing}

The electronic clustering or localization process described above provides an
average hole-hole attraction.\cite{PhysC12} This pair-potential is used 
in the Bogoliubov-deGennes (BdG) self-consistent calculations. 
Such calculations were done on simulated charge disordered systems of
$28 \times 28$ sites in a square lattice
with a hopping integral $t=0.15$eV. 
The $x=0$ Zn doping system is known to have an average 
$\Delta_d(x=0,T\approx 0) \sim 41$meV
from many experiments\cite{Hufner} and this is simulated by 
an attractive pair potential at $T=0$ of $2.30t$.

In the simulations, the phase separation is essentially completed 
with 40000 time steps and this is taken as equivalent to $x=0$.
As the Zn concentration increases, we assume the phase separation is diminished and
we decrease the pair-potential in the 
same proportion, as the free energy barrier depicted in Fig.(\ref{FEtimeEvol})
diminishes. With this procedure and the BdG calculations, we  obtain the $x$-dependent $\Delta_d(x,T\approx 0)$.
The results for zero and finite $x$ are summarized in Table.1

\section{Calculations of $T_c$ by Phase Coherence}

We now apply the formalism of a granular superconductor
and the assumption that the Zn atoms weaken the $d$-wave pair
potential to calculate the variation of $T_c(x)$. 
At low temperatures the patches form independent SC small grains interacting 
with one another via Josephson coupling,\cite{PhysC12} exactly
as a granular superconductor.\cite{Ketterson} 

\begin{figure}[ht]
\begin{center}

\centerline{\includegraphics[width=6.5cm,angle=-90]{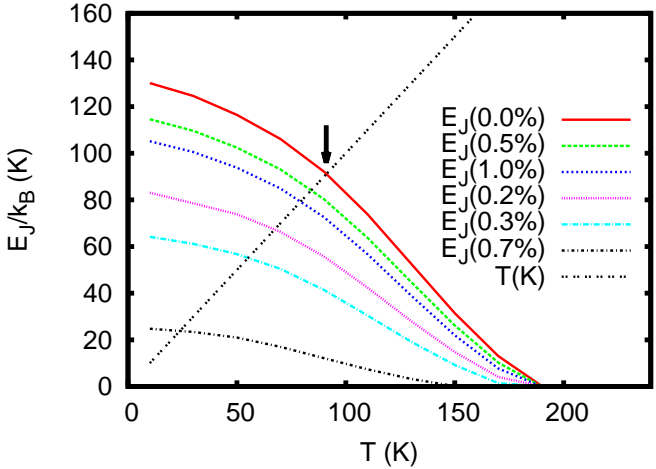}}
   \end{center}
\caption{(Color online) The estimation of $T_c(x)$ from Eq.(\ref{EJ})
with the calculated values of $\Delta_d^{Av}(T)$ and the the values
of the normal resistivity $R_n$ taken from the experimental 
$\rho_{ab}(p,T\gtrsim T_c)$. The relation between the time step and
the Zn percentage $x$ is listed in Table 1.
The arrow indicates $T \! = \! T_c$ for x=0, the undoped system. 
}
\label{Ejs} 
\end{figure}

Detailed studies of the $d$-wave weak links between superconductors
have shown that the tunneling currents depends mainly on the maximum
lobe amplitude and qualitatively resembles that of
a $s$-wave superconductor.\cite{Barash,Bruder} 
Under this approximation we assume the average Josephson coupling energy $E_J^{\rm Av}$ between the puddles to
be the simple analytical expression derived for coupling between two similar $s$-wave superconductors \cite{AB}
\begin{eqnarray} 
E_J^{\rm Av}(x,T) = \frac{\pi h\Delta^{\rm Av}_d(x,T)}{2 e^2 R_n(x)}
\tanh \left[\frac{\Delta^{\rm Av}_d(x,T)}{2k_BT} \right] \, ,
\label{EJ} 
\end{eqnarray}
where $\Delta^{\rm Av}_d(x,T) \! = \! \sum_i^N\Delta_d({\bf r}_i,x,T)/N$ is the average energy gap of all 
regions of Cooper pairing, and $R_n(x)$ is the average normal resistance between 
neighboring patches at a temperature just above the phase coherence temperature
$T_c(x)$. $R_n(x)$ is proportional to the normal state in-plane resistivity 
$\rho_{ab}(x,T\gtrsim T_c)$ just above $T_c(x)$ measured in the experiments.\cite{Xu,Marta,Walker}
As explained previously,
when the temperature is lowered, thermal fluctuations diminish, and long-range phase coherence 
is achieved when $k_BT \! \approx \! E_J^{\rm Av}(T)$ at $T_c$. The Josephson 
coupling of Eq.(\ref{EJ}) and $k_BT_c$ are plotted in Fig.(\ref{Ejs}).

The transport properties of Zn-doped materials were studied by many 
groups.\cite{Alloul,Xu,Marta,Walker} We take their values of $T_c(x)$ and use their 
data of $\rho_{ab}(x,T\gtrsim T_c)$ in Eq.(\ref{EJ}). We summarize the experimental
and theoretical results to optimum YBCO in Table 1.
\begin{table}[h]
\begin{tabular}{|l|l|l|l|l|l|}
\hline
Zn($\%$)& $T_c^{exp}$ & $\rho_{ab}$ & Step & $\Delta_d^{Av}(0)$ & $T_c^{theo}$ \\ \hline
0 & 91 K & 7.0 & 40000 & 42.8 meV & 91.0 K \\ \hline
0.5 & 84 K & 7.7 & 26000 & 41.5 meV & 84.0 K \\ \hline
1.0 & 79 K & 8.0 & 25730 & 39.4 meV & 79.0 K \\ \hline
2.0 & 67 K & 9.4 & 20000 & 36.4 meV & 67.2 K \\ \hline
3.0 & 56 K & 11.0 & 9500 & 33.2 meV & 55.0 K \\ \hline
7.0 & 25 K & 19.0 & 6000 & 22.0 meV & 24.0 K \\ \hline
\end{tabular}
\caption{ The first three columns are experimental results of optimum
YBCO $T_c$ from two experiments.\cite{Xu,Walker} $\rho_{ab}$ is 
in units of $(\mu\Omega m)$.
The last column has our calculations of $T_c$ using the results 
of $\Delta_d^{Av}(T)$ in Eq.(\ref{EJ}) and $T_{PS}\sim 300$K.}
\end{table}

\begin{figure}[ht]
\begin{center}
     \centerline{\includegraphics[width=9.0cm
     ,angle=0]{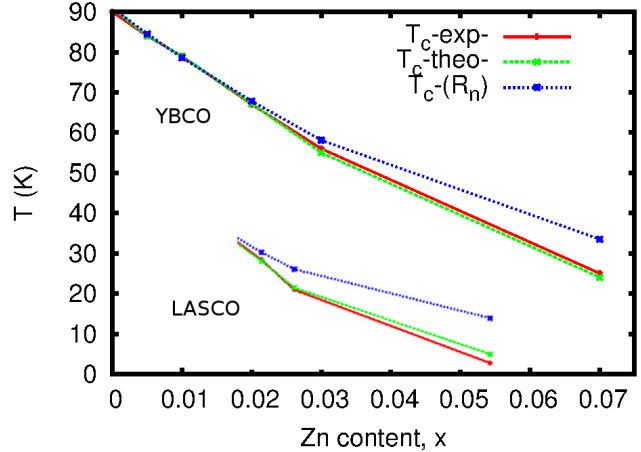}}
   \end{center}
\caption{(Color online) The experimental critical temperature~\cite{Xu,Walker} (red),
the calculations assuming only the normal resistivity (blue) $\rho_{ab}(p,T\gtrsim T_c)$
and the theoretical $T_c$ using Eq.(\ref{EJ}) (green) and the parameters listed in 
Table 1 for optimum YBCO.  At bottom we plot the same for LASCO for x=0.02, 0.025
and 0.055.
}
\label{Tcs} 
\end{figure}

In Fig.(\ref{Tcs}) we plot our main results; the theoretical values of $T_c(x)$ extracted from
Eq.(\ref{EJ}) and Fig.(\ref{Ejs})
with the experimental data listed in Table 1. We also plot a ``pair-breaking''
estimation, {\it i.e.}, $T_c(x)=T_c(0)\times \rho_{ab}(0)/\rho_{ab}(x)$. This linear
approximation works well only for $x<3\%$. At the bottom of Fig.(\ref{Ejs})
we show results for a film of LASCO with $T_c(x=0)=34$K with experimental
data from Ref. \cite{Marta} and similar calculations but with 
$T_{PS}\sim 200$K.

In our approach, the measured onset of Nernst signal\cite{Xu} $T_{co}(x)$
occurs due to the vortex motion between the patches with higher electronic
density that have the larger local $d$-wave amplitude. The Zn doping effect
on $T_{co}(x)$ is the same of that on $T_c$ because both depend on the 
superconducting properties.
Concerning the behavior of $T^*(x)$, we do not have a way
to calculate this temperature but, if its origin is related to the phase separation, 
recalling that we use only one single value of $T_{PS}\sim 300$K for Zn doped optimum YBCO, it
should be independent of $x$, as it is measured\cite{Alloul,Xu}. On the other
hand, if it depends on the superconducting properties it should also decrease, like
$T_c(x)$ and the onset of Nernst temperature $T_{co}(x)$.\cite{Xu,Walker,Abe} 
A more accurate way than resistivity measurements to resolve between these two 
possibilities is to perform TF-$\mu$SR, similar to what was done on cuprate,\cite{Mahyari} 
to determine if the onset of diamagnetic response depends on the Zn concentration.

\section{Conclusion}

We have proposed that cuprate superconductors undergo a phase separation
transition that leads to the formation of incommensurate charge disorder
breaking spatial invariance. In this scenario, the Fermi surface
rather than being globally defined, changes on nanometer length scales, 
as observed in some experiments.\cite{Lang,Wise,Neto}
Contrary to current ideas of competing orders, the charge segregation
transition favors the formation of local Cooper pairs and
may be the origin of the superconducting state.
The electronic regions of low and high density play the hole of grains
in a granular superconductor on which $T_c$ is reached by Josephson
coupling. This approach explains the STM tunneling asymmetry of cuprates,\cite{Neto}
and several important experiments 
reporting some sort of inhomogeneous responses. In particular,
we show here to be a key ingredient  to reproduce the Zn-doped data 
on different systems.
In the light of the recent incommensurate charge inhomogeneities found
in many different cuprate systems, our work furnishes
strong evidence of this universal behavior and provides also a 
simple interpretation to the pseudogap phase above $T_c$.

\acknowledgments
We acknowledges partial financial aid from the Brazilian agencies FAPERJ and CNPq
and CAPES.


\begin{thebibliography}{
0
}

 \bibitem{Alloul}
  \Name{Alloul H., Bobroff J.,Gabay M., \and Hirschfeld P.J.}
  \REVIEW{Rev. Mod. Phys.}{\bf 81}{45}{(2009)}.
  
 \bibitem{Bernhard}
  \Name{Bernhard C., Tallon J.L., Bucci C.,DeRenzi R., Guidi G., Williams G.V.M., Niedermayer Ch.}
  \REVIEW{Phys. Rev. Lett}{\bf 77}{2304}{(1996)}.

\bibitem{Xu}
  \Name{Xu Z.A., Shen J.Q., Zhao S.R., Zhang Y.J., \and Ong C.K.}
  \REVIEW{Phys. Rev. B}{\bf 72}{144527}{(2005)}.

\bibitem{Marta}
  \Name{Cieplak Marta Z., et. al.}
  \REVIEW{App. Phys. Lett.}{\bf 73}{2823}{(1998)}.

\bibitem{Walker}
  \Name{Walker D.J.C., Mackenzie A.P., \and Cooper J.R.}
  \REVIEW{Phys. Rev. B}{\bf 51}{15653}{(1995)}.

\bibitem{Abe}
  \Name{Abe Y., Segawa K., \and Ando Y.}
  \REVIEW{Phys. Rev. B}{\bf 60}{R15055}{(1999)}.

\bibitem{Lang}
  \Name{Madhavan V., Hoffman J.E., Hudson E.W., Eisaki H., 
Uchida S., \and Davis J.C.}
  \REVIEW{Nature}{\bf 415}{412}{(2002)}.

\bibitem{Gomes}
  \Name{Gomes Kenjiro K.,
Pasupathy Abhay N., Pushp Aakash, Ono Shimpei,
Ando Yoichi, \and Yazdani Ali}
  \REVIEW{Nature}{\bf 447}{569}{(2007)}.


\bibitem{McElroy}
  \Name{McElroy K., Lee D.H., Hoffman J. E., Lang K. M,
Hudson E. W., Eisaki H., Uchida S., Lee J., \and Davis J.C.}
  \REVIEW{Phys. Rev. Lett.}{\bf 94}{197005}{(2005)}.

\bibitem{Pasupathy}
  \Name{Pasupathy Abhay N., 
Gomes Kenjiro K., Parker Colin V., Wen Jinsheng, Xu Zhijun, Gu Genda, Ono Shimpei, Ando Yoichi, \and Yazdani Ali}
  \REVIEW{Science}{\bf 320}{196}{(2008)}.
%

\bibitem{Kato}
  \Name{Kato T., Maruyama T., Okitsu S., \and Sakata H.}
  \REVIEW{J. Phys. Soc. Jpn.}{\bf 77}{054710}{(2008)}.


\bibitem{Wise}
  \Name{Wise W. D., Boyer M. C., Chatterjee Kamalesh, Kondo Takeshi, Takeuchi T., Ikuta H., Wang Yayu \and Hudson E.W.}
  \REVIEW{Nature Physics}{\bf 4}{696}{(2008)}.

\bibitem{Ghiringhelli}
  \Name{Ghiringhelli G., Le Tacon M., Minola M., Blanco-Canosa S., Mazzoli C, Brookes N.B., De Luca G.M., Frano A., Hawthorn D.G., He F., Loew T., Moretti Sala M., Peets D.C., Salluzzo M., Sutarto R., Sawatzky G.A., Weschke E., Keimer B. \and Braicovich L.}
  \REVIEW{Science}{\bf 337}{821}{(2012)}.


\bibitem{Chang}
  \Name{Chang J., Blackburn E., Holmes A.T., Christensen N.B., Larsen J., Mesot J., Liang R., Bonn D.A., Hardy W.N., Watenphul A., Zimmermann M.V., Forgan E.M. \and Hayden S.M.}
  \REVIEW{Nature Physics}{\bf 8}{871}{(2012)}.

\bibitem{Torchinsky}
  \Name{Torchinsky D.H., Mahmood F., Bollinger A.T., Bo\`zovi\'c I.
\and Gedik N.}
  \REVIEW{Nature Materials}{\bf 12}{387}{(2014)}.


\bibitem{LeTacon}
  \Name{Le Tacon M., Bosak A., Souliou S.M., Dellea G., Loew T., Heid R., Bohnen K.P., Ghiringhelli G., 
Krisch K. \and Keimer B.}
  \REVIEW{Nature Physics}{\bf 10}{52}{(2014)}.


\bibitem{Comin}
  \Name{Frano A., Yee M.M., Yoshida Y., Eisaki H., Schierle E., Weschke E., Sutarto R., He F., Soumya-narayanan A., He Y., Le Tacon M., Elfimov I.S., Hoffman J.E., Sawatzky G.A., Keimer B. \and Damascelli A.}
  \REVIEW{Science}{\bf 343}{390}{(2014)}.


\bibitem{Neto}
  \Name{Aynajian P., Frano A., Comin R., Schierle E., Weschke E., Gyenis A., Wen J., Schneeloch J., Xu Z., Ono S., Gu G., Le Tacon M. \and Yazdani A.}
  \REVIEW{Science}{\bf 343}{393}{(2014)}.


\bibitem{Sonier:08}
  \Name{Sonier J.E., Ilton M., Pacradouni V., Kaiser C.V., Sabok-Sayr S.A., 
Ando Y., Komiya S., Hardy W.N., Bonn D.A., Liang R. \and Atkinson W.A.}
  \REVIEW{Phys. Rev. Lett.}{\bf 101}{117001}{(2008)}.


\bibitem{Mahyari}
  \Name{Mahyari Z.L., Cannell A., de Mello E.V.L., 
Ishikado M., Eisaki H., Liang R., Bonn D.A., Hardy W.N. \and Sonier J.E.}
  \REVIEW{Phys. Rev. B}{\bf 88}{144504}{(2013)}.


\bibitem{Mello04}
  \Name{de Mello E.V.L. \and Caixeiro E.S.}
  \REVIEW{Phys. Rev. B}{\bf 70}{224517}{(2004)}.

\bibitem{Mello09}
  \Name{de Mello E.V.L., Kasal R.B. \and Passos C.A.C.}
  \REVIEW{J. Phys. C.M.}{\bf 21}{235701}{(2009)}.

\bibitem{CH}
  \Name{Cahn J.W. \and Hilliard J.E.}
  \REVIEW{J. Chem. Phys.}{\bf 28}{258}{(1958)}.

\bibitem{Otton}
  \Name{de Mello E.V.L. \and Silveira Filho O.T.}
  \REVIEW{Physica A}{\bf 347}{429}{(2005)}.


\bibitem{PhysC12}
  \Name{de Mello E.V.L. \and Kasal R.B.}
  \REVIEW{Physica C}{\bf 472}{60}{(2012)}.

\bibitem{Harris}
  \Name{Brooks Harris A.}
  \REVIEW{J. Phys. C: Solid State}{\bf 7}{1671}{(1974)}.

\bibitem{Muller}
  \Name{Sigmund E. \and M\"uller K.A.}
  \REVIEW{"Phase Separation in Cuprate Superconductors"}{}{}{(1993)}.

\bibitem{Bray}
  \Name{Bray A.J.}
  \REVIEW{Adv. Phys.}{\bf 43}{347}{(1994)}.


%



%
%

\bibitem{EPL12a}
  \Name{de Mello E.V.L.}
  \REVIEW{Europhys. Lett.}{\bf 98}{57008}{(2012)}.

\bibitem{EPL12b}
  \Name{de Mello E.V.L.}
  \REVIEW{Europhys. Lett.}{\bf 99}{37003}{(2012)}.

\bibitem{EPL13}
  \Name{de Mello E.V.L. \and M\"ockli D.}
  \REVIEW{Europhys. Lett.}{\bf 102}{17008}{(2013)}.

\bibitem{Hufner}
  \Name{H\"ufer S., Hossain M.A., Damascelli A. \and Sawatzky G.A.}
  \REVIEW{Rep. Prog. Phys.}{\bf 71}{062501}{(2008)}.


\bibitem{Ketterson}
  \Name{Ketterson J.B. \and Song S.N.}
  \REVIEW{"Superconductivity"}{\bf }{}{(1999)}.

\bibitem{Barash}
  \Name{Barash Y.S., Galaktionov A.V. \and Zaikin A.D.}
  \REVIEW{Phys. Rev. B}{\bf 52}{665}{(1995)}.


\bibitem{Bruder}
  \Name{Bruder C., van Otterlo A. \and Zimanyi G.T.}
  \REVIEW{Phys. Rev. B}{\bf 51}{R12904}{(1995)}.


\bibitem{AB}
  \Name{Ambeogakar V. \and Baratoff A.}
  \REVIEW{Phys. Rev. Lett.}{\bf 10}{486}{(1963)}.


\end{thebibliography}
\end{document}